\documentclass[12pt,reqno,a4paper]{amsart}

\usepackage{lipsum,subeqnarray}
\usepackage{amsfonts}
\usepackage{graphicx}
\usepackage{amsmath}
\usepackage{mathtools}
\usepackage{amssymb}
\usepackage{enumerate}
\usepackage{slashed}
\usepackage{graphicx}
\usepackage{newlfont}
\usepackage{amsrefs}
\usepackage{comment}
\usepackage[normalem]{ulem}
\usepackage{mathtools}
\usepackage{accents}
\usepackage{leftidx}
\usepackage{bbm}
\usepackage{amsthm}

\numberwithin{equation}{section}

\theoremstyle{plain}
\newtheorem{theorem}{Theorem}[section]

\newtheorem{cor}{Corollary}

\newtheorem*{theorem*}{Conjecture}

\usepackage{xcolor}

\theoremstyle{definition}

\theoremstyle{remark}

\usepackage{stackengine}
\usepackage{subeqnarray} 

\stackMath
\newcommand\tenq[2][1]{%
 \def\useanchorwidth{T}%
  \ifnum#1>1%
    \stackunder[0pt]{\tenq[\numexpr#1-1\relax]{#2}}{\scriptscriptstyle\sim}%
  \else%
    \stackunder[1pt]{#2}{\scriptscriptstyle\sim}%
  \fi%
}

\begin{document}
\title[Galaxy rotation curves incorporating relativistic mass]{Galaxy rotation curves incorporating relativistic mass and comparison with Keplerian velocity curves}
 
\author{Jaroslaw S. Jaracz}

\begin{abstract}
We derive a formula for the velocity distribution of an axially symmetric galaxy where the mass density is corrected using the mass formula from special relativity. We take some reasonable test mass densities and numerically compute the resulting galaxy rotation curves. We then compare these to the rotation curves obtained from a similar formula without a relativistic correction factor. We find that the correction factor has a small dark-matter like effect. Finally, we compare these to the corresponding Keplerian velocity curves. We find that there is a large discrepancy in this case, where away from the galactic center, up to the galactic edge, the curves computed using our formulas give a noticeably higher velocity.
\end{abstract}
\maketitle

\section{Introduction}\label{SEC:Intro}

The problem of dark matter is well known. The rough idea is that the stars far away from the galactic center have velocities larger than they should. That is, based on the mass enclosed within the orbit of a star, where the mass calculated from luminosity data, the velocity is significantly larger than predicted by Kepler's law. One way of remedying the situation is to propose the existence of large quantities of as-of-yet unobserved matter which only interacts through gravity and doesn't have any luminosity properties. 

However, it still remains a mystery as to what exactly dark matter is. There have been many well designed experiments with the goal of detecting dark matter, but so far all have yielded null results. Moreover, if dark matter is so plentiful, and assuming the solar system is not a special location, it should be all around us. This means that either dark matter is just extremely weakly interacting, or there is less of it than predicted. Thus it is extremely important to determine exactly what the expected distribution of dark matter is in a galaxy. 

One good candidate for dark matter is black holes. Black holes have at this point been experimentally observed, with the now-famous photo of a black hole at the center of the Milky Way. They only interact through gravity, and classically they do not radiate since light can't escape from them. Quantum mechanically, even if Hawking radiation is a real phenomenon, then the temperature of black holes would be so low that the thermal radiation could not be detected against the cosmic microwave background. 

In fact, the primary problem with black holes as candidates for dark matter is that there would be too many of them. Based on current models of stellar evolution, the number of black holes should be much smaller than the required amount of dark matter to explain the galactic rotation curves. Also, there is no known mechanism in the early universe which would produce the required number of black holes.

For this reason, it is very important to determine exactly how much dark matter is actually needed to explain the galactic rotation curves. In order to perform the calculations, simplifying assumptions are often made, and certain effects neglected. However, in light of the many null experiments, it might be worthwhile to take these effects into account, and remove some of the simplifying assumptions. In this paper we only look at a few of these assumptions. However, there are several more that are worthwhile of further investigation at a later date, and so we do mention them. 

First off, the galactic rotation curves are calculated based on Kepler's second law. However, this law is derived from Newton's gravitational law for a point particle in orbit around another point particle. It also holds in the case the particles have spatial extent assuming they are spherically symmetric, via Newton's shell theorem. It is important to remember, Newton's shell theorem for a $1/r^2$ potential only works in dimension $n=3$. However, galaxies are not spherically symmetric, but rather disk shaped. So it doesn't make sense to expect Kepler's law to hold in such a case. It is easy to see why. Consider a spherically symmetric object of some fixed mass, choose some plane through the object, and consider a particle orbiting the object in that plane. What happens if the object is squished into a thin plate in the plane? The centripetal force on the orbiting particle is increased since all of the gravitational force on the particle is now acting in the plane, so the velocity must be increased to maintain the same radius for the orbit.

In fact, near the galactic core, which can be accurately approximated as spherically symmetric, the observed velocity curves do match the ones predicted from Kepler's law. It is only far away from the galactic center, as the spherical symmetry is lost, that the curves begin to diverge. Thus, in this paper we model our galaxy as a disk of finite but non-zero thickness.

Next, the effective mass of stars far away from the galactic center is not equal to the mass calculated from the luminosity data. Notice that since the stars are moving with some velocity, we should take their mass the be the rest mass times the special relativistic correction factor. This is of course a small effect. However, it is also cumulative in the following sense. Suppose that the correction is ignored for the mass enclosed by the orbit of a star. Then, this mass is underestimated. Therefore, the velocity of the star is underestimated as well. Therefore, its effective relativistic mass is underestimated, and hence the effect is compounded for any stars whose orbits enclose the orbit of the given star. The easiest way to determine the significance of this effect is to investigate it numerically. This is what is done in this paper. 

There are also at least two more relativistic effects which contribute to the underestimation of the mass. We do not investigate them in detail here, but we mention them to motivate further research in the future. The first is the issue of gravitational binding energy. As is well known (see \cite{Wald}, Chapter 6.2) the ADM mass of a spherically symmetric star is larger than just the integral of its mass density. This positive difference is known as the gravitational binding energy. The same effect occurs is gravitating systems with multiple components, and would increase the effective mass of the system. 

Finally, there is the issue of the red shift. Photons escaping a strong gravitational field are of course red shifted. This red shift would be larger for photons emitted in the dense core of the galaxy. Also, there is the contribution to the red shift from cosmological expansion. This red shift has no effect on the measured velocity of rotation. However, if it is not taken into account accurately, then it results in an underestimation of the temperature, and hence the mass of the star. Of course, photons entering the Milky Way are blue shifted. But, if one observes a galaxy much more massive than the Milky Way, there would be a net red shift. This effect might be worth investigating.

We also briefly mention the issue of the bullet cluster. Notice, the bullet cluster consists of two galaxies passing through each other. In the reference frame of one galaxy, the other one is moving at quite a high velocity. Hence, there is all the more reason to take a relativistic correction to the mass for the purpose of calculating the gravitational lensing effects. 

In isolation, each of these effects is small and can be argued as negligible. However, when all of them are taken into account, they might be enough to decrease the amount of dark matter needed to explain the rotation curves to the point where some alternate hypothesis, such as black holes, might give a reasonable explanation for the missing mass. In light of the many null detection experiments, this would be the simplest explanation. And, even if all of these effects do not account for all of the missing matter, they might suggest why it has been so difficult to detect the dark matter, simply because there is less of it than was thought. 

The result of this paper is mathematical in nature. We propose a model for an axially symmetric, stable galaxy. We then derive a formula for the velocity as a function of the radius and the axial coordinate, where the formula takes into account the special relativistic correction to the mass density. We then numerically apply the formula to some simple model mass densities (which experimentally could be obtained from luminosity data) and compare the resulting rotation curves to the case where the special relativistic correction is not taken into account. 

This gives a very small dark matter like effect, which only becomes apparent at very high galactic densities, and the effect becomes smaller as the density becomes small and physically reasonable. However, much more interesting is the comparison of these velocities to the corresponding Keplerian velocity. In that case we find that away from the galactic center, the velocity obtained from our formula where we ignore the relativistic correction, is noticeably larger than the Keplerian velocity. Moreover, this effect persists even for small densities, meaning that if we scale our mass density by some small parameter $\alpha$ to make it physically more reasonable, the ratio of the two velocities remains constant. See the figures in Section 5.4.

The results are very intriguing and suggest that the next logical step is to obtain the radial density functions of some real galaxies using luminosity data, plug these into the model, and compare the resulting velocities to the experimentally observed rotation curves.

\section{Statement of Results}

We work in cylindrical spatial coordinates $(r, \theta, z)$ and we use units so that $c=G=1$. We assume each star travels in some circular orbit in the plane $z=C$ for some $C$. We approximate our galaxy as a continuum. In that case, our galaxy can be described by its mass density $\rho=\rho(r, \theta, z, t)$ and velocity distribution $v=v(r, \theta, z, t)$. We assume that the galaxy is stable and axially symmetric, meaning that functions do not depend on $t$ or $\theta$. We also assume the galaxy is symmetric with respect to the plane $z=0$. Moreover, we make the approximation that the velocity is constant in $z$. This is an acceptable approximation based on observation. Thus we have $\rho=\rho(r, z)$ and $v=v(r)$. We assume that the static mass distribution is given by $\rho_0=\rho_0 (r, z)$. This is the mass distribution if the galaxy was not rotating. In practice, this would be obtained by taking a picture of the galaxy and extrapolating the mass density from the luminosity data. We assume that our system follows Newton's law of gravitation, except that the ordinary Newtonian mass is replaced by its special relativistic value. Thus we have
\begin{equation}
    \rho(r, z) = \frac{\rho_{0}(r, z)}{\sqrt{1-v^2(r)}}.
\end{equation}
Under these circumstances we have the following theorem.

\begin{theorem} \label{MainTheorem1}
The velocity distribution for an axially symmetric stable galaxy satisfying the assumptions above is given by the integral equation
\begin{align} \label{MainFormula}
    \frac{v^2(R)}{R} =\int_0^\infty \int_{0}^{2\pi} \int_{-\infty}^{\infty}  \frac{\rho_{0}(r, z)}{\sqrt{1-v^2(r)}} \frac{R-r\cos(\theta)}{d^{3}} r dz dr d\theta
\end{align}
where $d=\sqrt{ (r\cos(\theta)-R)^2 + r^2 \sin^2(\theta) + z^2  }$. 
\end{theorem}

\begin{cor}
Under the assumptions of Theorem \ref{MainTheorem1}, but ignoring the relativistic effects, the velocity is given by
\begin{align} \label{MainFormula2}
    \frac{v^2(R)}{R} =\int_0^\infty \int_{0}^{2\pi} \int_{-\infty}^{\infty}  \rho_{0}(r, z) \frac{R-r\cos(\theta)}{d^{3}} r dz dr d\theta
\end{align}
where $d=\sqrt{ (r\cos(\theta)-R)^2 + r^2 \sin^2(\theta) + z^2  }$. 
\end{cor}

Notice, one has $V(0)=0$. For a given static mass distribution $\rho_0$, the integral equation \eqref{MainFormula} can be numerically solved via Picard iteration. That is, one starts with $v_0=0$ and then defines
\begin{align} \label{IterativeFormula}
    \frac{v_{n+1}^2(R)}{R} =\int_0^\infty \int_{0}^{2\pi} \int_{-\infty}^{\infty}  \frac{\rho_{0}(r, z)}{\sqrt{1-v_n^2(r)}} \frac{R-r\cos(\theta)}{d^{3}} r dz dr d\theta
\end{align}
where in the usual way, for a well behaved $\rho_0$, the sequence of functions converges. The $\rho$ has to be sufficiently small. From a physical point of view this makes sense, for if the density was too large, such a galaxy wouldn't be even approximately stable. Thus, \eqref{MainFormula} is easily tractable numerically. 

\section{Proof of Theorem \ref{MainTheorem1}}

The proof is just basic calculus. We take cylindrical coordinates $(r, \theta, z)$. Since we are making the approximation that $v=v(r)$ and that we are working in axial symmetry, we only have to calculate the velocity at the point $(R, 0, 0)$. We will do so by calculating the magnitude of the centripetal acceleration which at $(R, 0,0)$ only has a component along the $x$-axis. 

Subdivide the $r$ interval into equal segments of length $\Delta r$, the $z$ interval into equal segments of length $\Delta z$, and the $\theta$ interval into equal pieces of length $\Delta \theta=2\pi/n$ so that we have the grid points $(i\Delta r, j \Delta \theta, k \Delta z)$ with $i\in \mathbb{Z}_{+}$, $ 1\leq j \leq n$, and $k\in \mathbb{Z}$. 

Using $(i, j, k)$ we can of course uniquely pick out a volume element $\Omega_{ijk}$ which contains the point $(r_i, \theta_j, z_k)=(i\Delta r, j \Delta \theta,  k\Delta z)$. The mass of such a volume element would then approximately be
\begin{equation*}
    m_{ijk} = \frac{\rho_0(r_i, z_k)}{\sqrt{1-v^2({r_i})}}r_i \Delta r \Delta \theta \Delta z
\end{equation*}
having taken into account the special relativistic correction factor. 

Take any $R$ and take some $r_m$ such that $|R-r_m|$ is minimal. If there is more than one such $m$, just take the smaller one. Let us calculate the force on $\Omega_{m00}$ due to $\Omega_{ijk}$ with $(i, j, k)\neq (m, 0, 0)$. We assume Newton's law applies. Converting to Euclidean coordinates, the distance between $(r_i, \theta_j, z_k)$ and $(r_m, 0, 0)$ is 
\begin{equation*}
    d^2=(r_i \cos(\theta_j) - r_m)^2 + r_i^2 \sin^2(\theta_j) + z_k^2
\end{equation*}
and the unit vector pointing from $\Omega_{m00}$ to $\Omega_{ijk}$ is 
\begin{align*}
    \hat{r} = \frac{1}{d}\left( r_i \cos(\theta_j) - r_m, r_i \sin(\theta_j), z_k  \right).
\end{align*}
By symmetry, there will not be any net acceleration on $\Omega_{m00}$ in either the $y$ or $z$ directions. Hence we are only interested in the $x$-component of the acceleration. In that case, applying Newton's law of gravitation, we get that the acceleration in the $x$-direction of $\Omega_{m00}$ due to $\Omega_{ijk}$ is 
\begin{align*}
    a_{x, ijk} = \frac{m_{ijk}}{d^2} \hat{r}_x =  \left(   \frac{\rho_0(r_i, z_k)}{\sqrt{1-v^2({r_i})}}r_i \Delta r \Delta \theta \Delta z \right) \frac{r_i \cos(\theta_j) - r_m}{d^3}.
\end{align*}
Summing up over all $\Omega_{ijk}$ except $\Omega_{m00}$, the net acceleration in the $x$-direction is 
\begin{align*}
    a_x = \sum_{\substack{i, j, k \\
    (i, j, k)\neq (m, 0, 0)}}     \frac{\rho_0(r_i, z_k)}{\sqrt{1-v^2({r_i})}} \frac{(r_i \cos(\theta_j) - r_m)}{( (r_i \cos(\theta_j) - r_m)^2 + r_i^2 \sin^2(\theta_j) + z_k^2    )^{3/2}} r_i \Delta r \Delta \theta \Delta z.
\end{align*}
At $(r_m, 0, 0)$ the centripetal acceleration is $(-a, 0, 0)=(a_x, 0, 0)$ where $a=v^2(r_m)/r_m$.  Therefore
\begin{align*}
    \frac{v^2(r_m)}{r_m} =  \sum_{\substack{i, j, k \\
    (i, j, k)\neq (m, 0, 0)}}     \frac{\rho_0(r_i, z_k)}{\sqrt{1-v^2({r_i})}} \frac{(r_m- r_i \cos(\theta_j))  }{( (r_i \cos(\theta_j) - r_m)^2 + r_i^2 \sin^2(\theta_j) + z_k^2    )^{3/2}} r_i \Delta r \Delta \theta \Delta z.
\end{align*}
Taking the limit as $\Delta r, \Delta \theta, \Delta z \rightarrow 0$, we have $r_m \rightarrow R$ and 
\begin{align*}
     \frac{v^2(R)}{R} =\int_0^\infty \int_{0}^{2\pi} \int_{-\infty}^{\infty}  \frac{\rho_{0}(r, z)}{\sqrt{1-v^2(r)}} \frac{R-r\cos(\theta)}{ ((r\cos(\theta)-R)^2 + r^2 \sin^2(\theta) + z^2)^{3/2}    } r dz dr d\theta
\end{align*}
as desired, and the corollary follows.

\section{Keplarian Velocity Curves}

Consider the mass enclosed in the orbit of radius $R$. The total mass, which we denote $M_K(R)$ is then
\begin{equation*}
    M_K(R)=\int_{-\infty}^{\infty} \int_0^{2\pi}  \int_0^R \rho_0(r, z) r dr d\theta dz = 2 \pi \int_{-\infty}^{\infty} \int_0^R \rho_0(r, z) r dr dz.
\end{equation*}
If this mass was spherically distributed (or concentrated at a point), then for a particle orbiting the spherical mass in an orbit of radius $R$, one would have
\begin{equation*}
    \frac{v^2}{R} = a = \frac{M_K(R)}{R^2}
\end{equation*}
so we have the Keplerian velocity distribution
\begin{align} \label{KeplerianVelocity}
    v_K(R)=\sqrt{\frac{M_{K}(R)}{R}}
\end{align}
which is reasonable to use as comparison for \eqref{MainFormula} and \eqref{MainFormula2}. Notice, assuming a spherical distribution of mass, we can ignore the effect of the mass density with $r>R$ by Newton's shell theorem.

\section{Numerical Analysis of Some Model Density Distributions}

A typical galaxy has a radius of about 50,000 light years and a thickness of 4,000 lightyears. Thus, scaling, we take $\rho_0(r, z)$ with compact support
\begin{equation*}
    \text{spt}  (\rho_0(r, z)) \subset [0, 5]\times [-0.2, 0.2].
\end{equation*}
For simplicity we take $\rho_0(r, z)=f(r)\Phi(z)$ for some function $f(r)$ and
\begin{equation*}
    \Phi(z) = \begin{dcases}
1 &: -0.2 \leq z \leq 0.2 \\
0 &: z<-0.2 \quad \text{or} \quad z>0.2
    \end{dcases}
\end{equation*}
so that the mass density is constant across the galactic disk and then vanishes outside of it. We take some simple functions for $f(r)$ and then numerically calculate the various velocities. 

We denote by $v_S$ the velocity given by \eqref{MainFormula}, where $S$ denotes the "Special relativistic velocity". We denote by $v_{NS}$ the velocity given by \eqref{MainFormula2}, where $NS$ stands for "Non-Special relativistic velocity". As before, we denote by $v_K$ the corresponding Keplerian velocity.

We mention that from numerical experiments, the difference between $v_S$ and $v_{NS}$ is very small for physically reasonable (that is, small) mass densities. To accentuate the effect, we take very large mass densities which lead to large velocities (recall in our units we have for the speed of light $c=1$), to show the separation in the curves. Taking a smaller mass density causes $v_S$ and $v_{NS}$ to get closer and closer together. 

However, this is not the case when comparing $v_{NS}$ and $v_K$. Their relationship remains constant. What we mean by this is if we take a mass density $\rho_0$ and scale it by some small factor $\alpha>0$ to obtain a new mass density $\tilde{\rho}_0 = \alpha \rho_0$ which is more physically reasonable, then we have
\begin{align*}
    \tilde{v}_{NS} = \sqrt{\alpha} v_{NS}, \quad \tilde{v}_K = \sqrt{\alpha} v_K
\end{align*}
and so the ratio of the two curves remains the same. 

In the following examples we let the number of Picard iterations be $N=3$, we take $N1=50$ points on the interval $[0, 5]$, $N2=50$ points on $[0, 2\pi]$, and $N3=21$ points on the interval $[-0.2, 0.2]$.  

\subsection{Exponentially decaying density, $f(r)=0.1e^{-r}$}

In this case we can explicitly calculate $M_K(R)$. We obtain
\begin{align*}
      M_K(R) &= 2 \pi \int_{-\infty}^{\infty} \int_0^R \rho_0(r, z) r dr dz = 2 \pi \int_{-\infty}^{\infty} \int_0^R \Phi(z)f(r) r dr dz \\
      & = 2 \pi \int_{-0.2}^{0.2} \int_0^R 0.1 e^{-r} r dr dz = (2\pi) (0.4) (0.1)  \int_0^R e^{-r} r dr \\
      &=\frac{2 \pi}{25} \left( 1-e^{-R}(R+1)  \right)
\end{align*}
so the Keplerian velocity in this case is
\begin{align*}
    v_{K}(R) = \sqrt{ \frac{2 \pi}{25R} \left( 1-e^{-R}(R+1) \right)   }.
\end{align*}

\subsection{Linearly decaying density, $f(r)=0.1(1-0.2r)$} 

We take \begin{equation*}
    f(r)=\begin{dcases} 0.1(1-0.2r) &: 0\leq r\leq 5 \\
    0 &: r>5 \end{dcases}.
\end{equation*}
In this case 
\begin{align*}
    M_K(R)=(2\pi)(0.4)(0.1) \int_0^R (1-0.2r) r dr = \frac{2 \pi}{25}\left(\frac{R^2}{2} - \frac{R^3}{15}\right)
\end{align*}
and so
\begin{equation*}
    v_k(R)=\sqrt{ \frac{2 \pi}{25R}\left(\frac{R^2}{2} - \frac{R^3}{15}\right)   } = \sqrt{  \frac{2 \pi}{25}\left(\frac{R}{2} - \frac{R^2}{15}\right)    }.
\end{equation*}

\subsection{Density decaying as $O(r^{-2})$ with $f(r)=0.1/(1+r^2)$}

With $f(r)=0.1/(1+r^2)$ we obtain
\begin{align*}
M_{K}(r) = (2\pi)(0.4)(0.1) \int_{0}^{R} \frac{1}{1+r^2} r dr = \frac{\pi}{25} \ln(1+R^2)   
\end{align*}
and so 
\begin{align*}
    v_K(r) = \sqrt{\frac{\pi}{25R} \ln(1+R^2)}.
\end{align*}

\subsection{Velocity Curves} In the three cases listed above we obtain the following graphs for the velocities.

\begin{figure}[h!]
\centering
\includegraphics[width=4.8in]{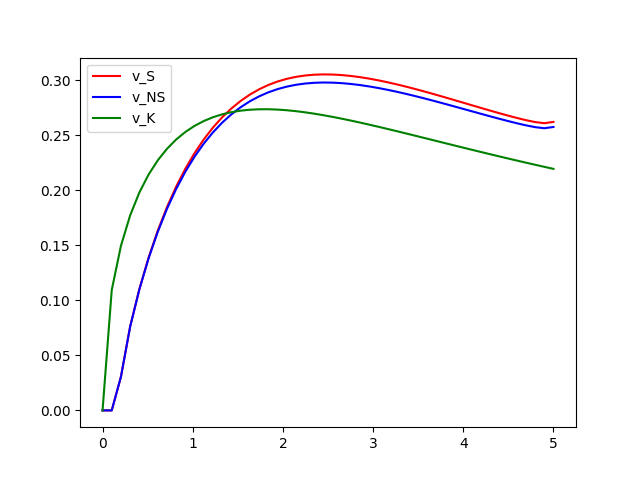}
\caption{Graphs of $v_S, v_{NS}$, and $v_K$ for $f(r)=0.1 e^{-r}$}
\end{figure}

\begin{figure}[h!]
\centering
\includegraphics[width=4.8in]{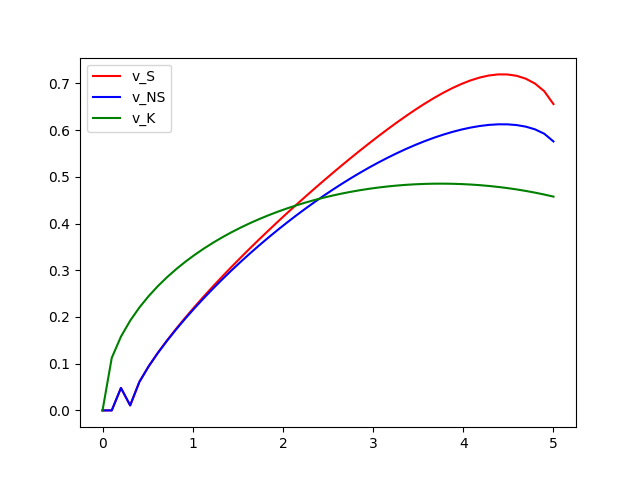}
\caption{Graphs of $v_S, v_{NS}$, and $v_K$ for $f(r)=0.1 (1-0.2r)$}
\end{figure}

\begin{figure}[h!]
\centering
\includegraphics[width=4.8in]{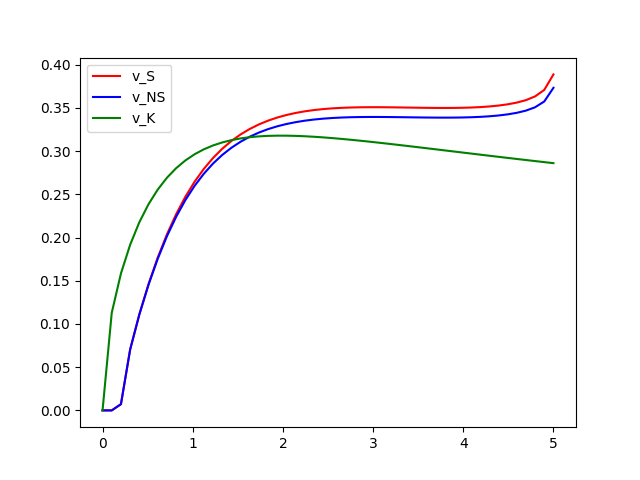}
\caption{Graphs of $v_S, v_{NS}$, and $v_K$ for $f(r)=0.1/(1+r^2)$}
\end{figure}

\section{Conclusion and Further Research}

We derived formula \eqref{MainFormula} for the velocity distribution $v_S$ of a stable, axisymmetric galaxy under some reasonable assumptions, taking into account the mass formula from special relativity. Ignoring this special relativistic correction, we obtained the velocity $v_{NS}$ given by \eqref{MainFormula2}. Due to the relativistic correction factor, $v_S>v_{NS}$ giving a small dark-matter like effect. Numerical experiments indicate that the effect is only noticeable for very large, unphysical densities. However, more interestingly, we then compared the resulting velocity curves to the Keplerian velocity $v_K$, where $v_K(R)$ was calculated assuming the mass enclosed by the orbit of radius $R$ was spherically distributed. 

As can be seen from the figures, for the given examples of centrally concentrated mass distributions, away from the galactic core, we have that $v_{NS}>v_K$ by a fairly noticeable amount. As mentioned earlier, the ratio $v_{NS}/v_{K}$ is unaffected by scaling the mass density by some small parameter $\alpha$, so the relationship for these curves continues to hold for small, physically reasonable mass distributions.

Naturally, this suggests applying the model to the density distributions of real galaxies based on luminosity data and comparing them to the observed galaxy rotation curves, which should be quite easy

\bibliographystyle{model1-num-names}
\bibliography{ref_new}

 \footnotesize

  J.S.~Jaracz, \textsc{Department of Mathematics, Texas State University,
    San Marcos, TX 78666}\par\nopagebreak
  \textit{E-mail address} \texttt{jsj74@txstate.edu}

\end{document}